\documentclass{article}
\usepackage{amsmath}
\usepackage{graphicx}

\begin{document}

\title{A model of radiating black hole in noncommutative geometry }
\author{Piero Nicolini \\
Dipartimento di Matematica e Informatica, Universit\`{a} di Trieste\\
Institut Jo\v{z}ef Stefan, Ljubljana\\
Dipartimento di Matematica, Politecnico di Torino, Turin\\
Istituto Nazionale di Fisica Nucleare, Sezione di Trieste}
\maketitle

\begin{abstract}
The phenomenology of a radiating Schwarzschild black hole is analyzed in a
noncommutative spacetime. It is shown that noncommutativity does not depend
on the intensity of the curvature. Thus we legitimately introduce
noncommutativity in the weak field limit by a coordinate coherent state
approach. The new interesting results are the following: i) the existence of
a minimal non-zero mass to which black hole can shrink; ii) a finite maximum
temperature that the black hole can reach before cooling down to absolute
zero; iii) the absence of any curvature singularity. The proposed scenario
offers a possible solution to conventional difficulties when describing
terminal phase of black hole evaporation.
\end{abstract}

\bigskip

In 1975 Hawking showed that a black hole is able to emit radiation and thus
to evaporate \cite{hawking}. This is one of most intriguing phenomenon in
the theory of gravitation, which, after 30 years still remains under debate
in particular for what concerns the mysterious explosive end of radiating
black holes (~see \cite{paddy} for a recent review with an extensive
reference list~). The black hole phenomenology is part of a larger research
area, whose final goal is the formulation of a full quantum theory of
gravity. In spite of the promising results that string theory has had in
quantizing gravity, the actual calculations of the Hawking radiation are
currently obtained by means of quantum field theory in curved space \cite
{swave}. In fact the black hole evaporation occurs in a semiclassical
regime, namely when the density of gravitons is lower than that of the
matter field quanta. In spite of this achievements the divergent behavior of
the black hole temperature in the final stage of the evaporation remains
rather obscure. Indeed in this extreme regime stringy effects cannot be
neglected. Recently an improved version of field theory on a noncommutative
space time manifold has been proposed as a cheaper way to reproduce the
stringy phenomenology, at least in the low energy limit. Noncommutativity is
encoded in the commutator
\begin{equation}
\left[ \,\mathbf{x}^{\mu }\ ,\mathbf{x}^{\nu }\,\right] =i\,\theta ^{\mu \nu
}.  \label{ncx}
\end{equation}
where $\theta ^{\mu \nu }$ is an anti-symmetric matrix which
determines a fundamental cell discretization of spacetime much in
the same way as the Planck constant $\hbar $ discretizes the phase
space. The basic motivation for assuming (\ref{ncx}) consists in
an attempt of curing the bad short distance behavior of pointlike
sources in ordinary field theory \cite{Nicolini 2004}, a problem
that is only partially solved even in the context of string
theory. Indeed preliminary results in a $2D$ noncommutative
spacetime suggest further investigation in this field
\cite{Nicolini 2005}.

The purpose of the paper is to understand whether noncommutativity
is able to cure the pathologies which occur in the theory of
gravitation, such as the divergent behavior of the black hole
temperature during the final stage of the collapse and the
curvature singularity at the black hole center. The most direct
way to reach this goal is to employ the linearized gravitational
field equations as a temporary laboratory to test the effect of
noncommutativity until the complete Einstein field equation will
be analyzed. There are three good reasons for doing this. First,
noncommutativity is an intrinsic property of the manifold and does
not depend on the curvature. Indeed there are many examples of
noncommutative field theory in flat space, namely in
\textit{absence of gravity}. Thus if any effect is produced by
noncommutativity it must appear also in the weak field regime.
Second, the concept of \textquotedblleft weak\textquotedblright\
or \textquotedblleft strong\textquotedblright\ field makes sense
only if one compares the field strength with a proper scale. In
the theory of gravitation we are given of a natural and unique
scale, that is the Planck scale. Thus with respect to the Planck
scale the gravitational field strength can still be considered
\textquotedblleft weak\textquotedblright\ even in the vicinity of
a black hole, justifying the adoption of linearized field
equations until the horizon radius is larger than the Planck
length. Third, we will show that, in the considered case of
Schwarzschild geometry, the expression of the temperature does not
depend on the weak field expansion.

In this spirit we assume as infinitesimal dimensionless parameter
\begin{equation}
\frac{\phi_N}{\phi_{Pl.}}\ll 1
\end{equation}
where $\phi_N$ is the Newtonian potential, while $\phi_{Pl.}=M_{Pl.}/l_{Pl.}$%
. As a peculiar aspects of Schwarzschild space time we observe that $g_{rr}$
is the only component of the metric tensor that is affected by
linearization. Indeed the line element reads
\begin{equation}
ds^2=-\left(1+\frac{2\phi_N}{\phi_{Pl.}}\right)dt^2 + dl^2
\end{equation}
where $dl^2$ is the Euclidean spatial line element. The Hawking temperature
is given by definition
\begin{equation}
T_H\equiv -\left(\,\frac{1}{4\pi\,\sqrt{-g_{00} \, g_{11}}}\, \frac{d g_{00}%
}{dr}\right)_{r=r_H}.
\end{equation}
As anticipated one finds the temperature $T_H=1/8\pi M$, that is coincident
with the expression given by the exact theory, since $T_H$ essentially
depends on $g_{00}$ only.

There are many approaches to implement noncommutativity in a field theory.
The underlying philosophy of these approaches is to modify the distribution
of point like sources in favor of smeared objects. Such prescription is in
agreement with the conventional procedure for the regularization of UV
divergences by the introduction of a cut off. In recent papers \cite{ae0,ae2}
the coordinate coherent state approach has been proposed and the precise
distribution of field sources has been determined to be the Gaussian
distribution. The resulting field theory is UV finite while Lorentz
invariance and unitariety are preserved \cite{ae}. In order to incorporate
noncommutativity one can observe that there exists a unique linearized
Einstein equation
\begin{equation}
\vec\nabla^2\, \phi_N = 4\pi\,G_N \,\delta\left(\, r\,\right),
\label{poiscl}
\end{equation}
namely the classical Poisson equation for a point-like source described by
Dirac delta-function, with $M$ the mass of the source and $G_N$ the Newton
constant. At this point we stress that noncommutativity is an \textit{%
intrinsic property} of the manifold, the contrary of curvature, that is a
geometrical tool defined over the underlying manifold to measure the
strength of the gravitational field. For these reasons curvature and
noncommutativity are independent concepts. Thus, noncommutative modification
of Schwarzschild space time, once introduced at a given curvature, will
remain valid in any other field strength regime. Our way of reasoning is in
perfect agreement with the standard procedure to implement noncommutativity
in quantum field theory where the strength of the field is not an issue \cite
{ae0}, \cite{ae2}. The physical effect of noncommutativity is that the very
concept of point-like object is no more meaningful and one has to deal with
\textit{smeared} objects only. To practical purpose the implementation of
noncommutativity by means of coherent state approach is realized by
substituting the position Dirac-delta, characterizing point-like structures,
with Gaussian function of minimal width $\sqrt{\theta}$ describing the
corresponding smeared structures. At this point the noncommutative field
equation reads \cite{ag}
\begin{eqnarray}
&&\vec\nabla^2\, \phi_N = 4\pi\,G_N \,\rho_\theta\left(\,\vec{x}\,\right)\ ,
\notag \\
&&\rho_\theta\left(\,\vec{x}\,\right)= \frac{M}{\left(\,2\pi\theta\,%
\right)^{3/2}}\, \exp\left(-\vec{x}^{\,2}/4\theta\,\right)  \label{poisson}
\end{eqnarray}
where $\rho_\theta$ is the Gaussian mass density. We obtain the NC version
of linearized Schwarzschild line element as
\begin{equation}
ds^2 = -\left( 1+ \frac{2\gamma \phi_N}{\sqrt{\pi} \phi_{Pl.}} \right) dt^2
+ dl^2  \label{ncs}
\end{equation}
where $\gamma$ is the lower incomplete Gamma function, with the definition
\begin{equation}
\gamma\equiv\gamma\left(1/2\ , r^2/4\theta\, \right)\equiv
\int_0^{r^2/4\theta} dt\, t^{-1/2} e^{-t}
\end{equation}
In the commutative limit $r/2\sqrt{\theta}\to\infty $ the standard
linearized Schwarzschild metric is reproduced. The line element (\ref{ncs})
describes the geometry of a noncommutative black hole under the condition
that $r_H\gg l_{Pl.}$ and should give us useful indications about possible
noncommutative effects on the Hawking radiation.

To calculate the Hawking temperature we need the event horizon radius $r_H$,
that is defined by the vanishing of ``unperturbed'' $g_{00}$. In our case it
leads to the implicit equation\footnote{%
We use convenient units $G_N=1$, $c=1$.}
\begin{equation}
r_H= \frac{2M}{\sqrt{\pi}}\,\gamma\left(1/2\ , r^2_H/4\theta\, \right)
\label{horizon}
\end{equation}
Rewriting (\ref{horizon}) in terms of the upper incomplete Gamma function as
\begin{equation}
r_H= 2M\,\left[\, 1 -\frac{1}{\sqrt\pi}\, \Gamma\left(1/2\ , r^2_H/4\theta\,
\right) \,\right]  \label{horizon2}
\end{equation}
one recovers the conventional Schwarzschild radius plus $\theta$%
-corrections. 
In the ``large radius'' regime $r^2_H/4\theta>>1$ equation (\ref{horizon2})
can be solved by iteration. At the first order approximation, we find
\begin{equation}
r_H= 2M\,\left(\, 1 - \sqrt{\frac{\theta}{\pi}}\frac{1}{M}\,
e^{-M^2/\theta}\,\right)
\end{equation}
The effect of noncommutativity is exponentially small, which is reasonable
to expect since for large radii with respect to $\sqrt{\theta}$, spacetime
can be considered as a smooth classical manifold (commutative limit). On the
other hand, in the opposite limit, one expects significant changes due to
the spacetime fuzziness. To this purpose it is convenient to invert (\ref
{horizon2}) and consider the black hole mass $M$ as function of $r_H$
\begin{figure}[h]
\begin{center}
\includegraphics[width=8cm,angle=270]{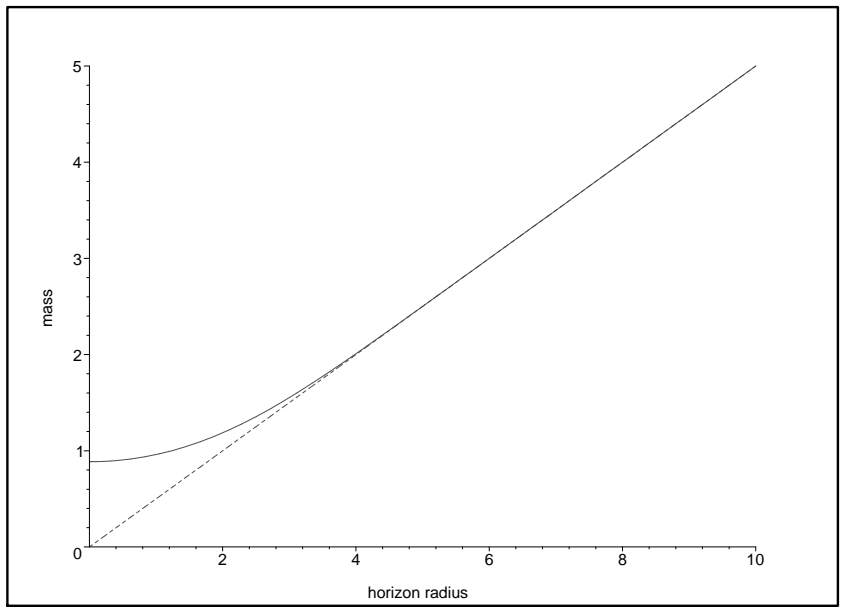}
\end{center}
\caption{Mass vs horizon relation. In the commutative case, dashed line, the
mass is the linear function $M=r_H/2$ vanishing at the origin, while in the
noncommutative case, solid line, $M\left(\, r_H\to 0\,\right)=M_0$, i.e. for
$M< M_0$ there is no event horizon.}
\label{fig1}
\end{figure}
\begin{equation}
M\left(r_H\right)= \frac{r_H\sqrt{\pi}}{2\gamma\left(1/2\ , r^2_H/4\theta\,
\right)}  \label{mh}
\end{equation}
In such a limit, i.e. $r_H/\sqrt{\theta}\ll 1$, the equation (\ref{mh})
leads to
\begin{equation}
M\rightarrow M_0=0.5\, \sqrt{\pi\,\theta}  \label{m0}
\end{equation}
which is a new and interesting result. Noncommutativity implies a minimal
non-zero mass that allows the existence of an event horizon (see Figure \ref
{fig1}). If the starting black hole mass is such that $M> M_0$, it can
radiate through the Hawking process until the value $M_0$ is reached. At
this point the horizon has totally evaporated leaving behind a massive
relic. Black holes with mass $M< M_0$ do not exist. An equivalent scenario
arises from the behavior of $g_{00}$ (see Figure \ref{fg00}). There are
three pictures

\begin{enumerate}
\item  for $M> M_0$ (dotted curve) there is a black hole with regular metric
in the origin;

\item  for $M= M_0$ (solid curve) the event horizon is shrank to the origin.

\item  for $M< M_0$ (dashed curve) there is no horizon.
\end{enumerate}

\begin{figure}[h]
\begin{center}
\includegraphics[width=8cm,angle=270]{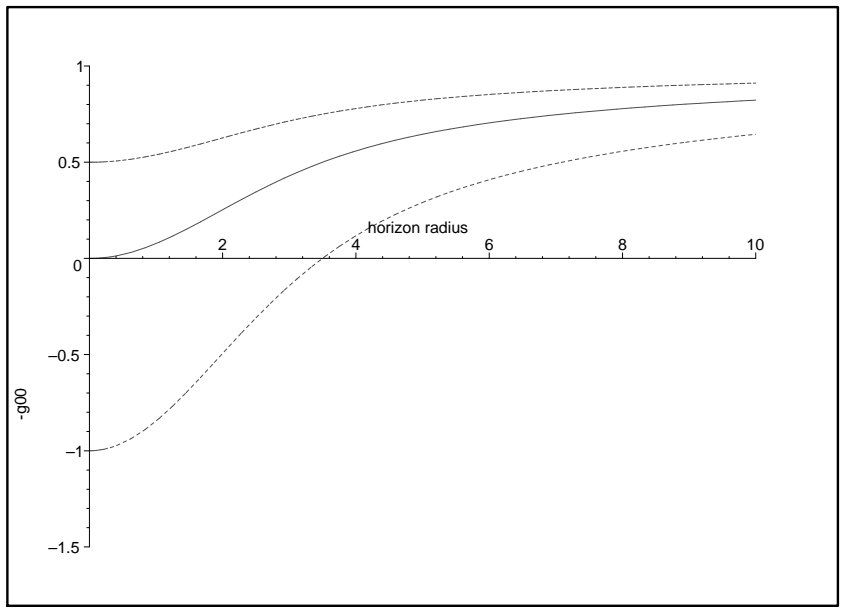}
\end{center}
\caption{The function $-g_{00}$ vs the radial distance $r$ for some value of
the mass $M$. The dashed line, corresponds to a mass $M=0.5 M_0$ for which
there is no event horizon. The dotted line corresponds to a mass $M=2 M_0$,
which describes a black hole, which is regular at its center $r=0$. The
solid curve is the borderline case, namely the case of $M=M_0$ in which the
horizon radius $r_H$ is shrank in the origin.}
\label{fg00}
\end{figure}

To understand the physical nature of the mass $M_0$ remnant, let us also
consider the black hole temperature as a function of $r_H$. It is given by
\begin{equation}
T_H\left(\, r_H \,\right)= \frac{1}{4\pi}\left[\, \frac{1}{r_H}-\frac{%
\gamma^\prime\left(\, 1/2\ ; r_H^2/4\theta\,\right)} {\gamma\left(\, 1/2\ ;
r_H^2/4\theta\,\right)}\, \right]  \label{thnc}
\end{equation}
where the ``prime'' denotes differentiation with respect to $r$.
\begin{figure}[h]
\begin{center}
\includegraphics[width=8cm, angle=270 ]{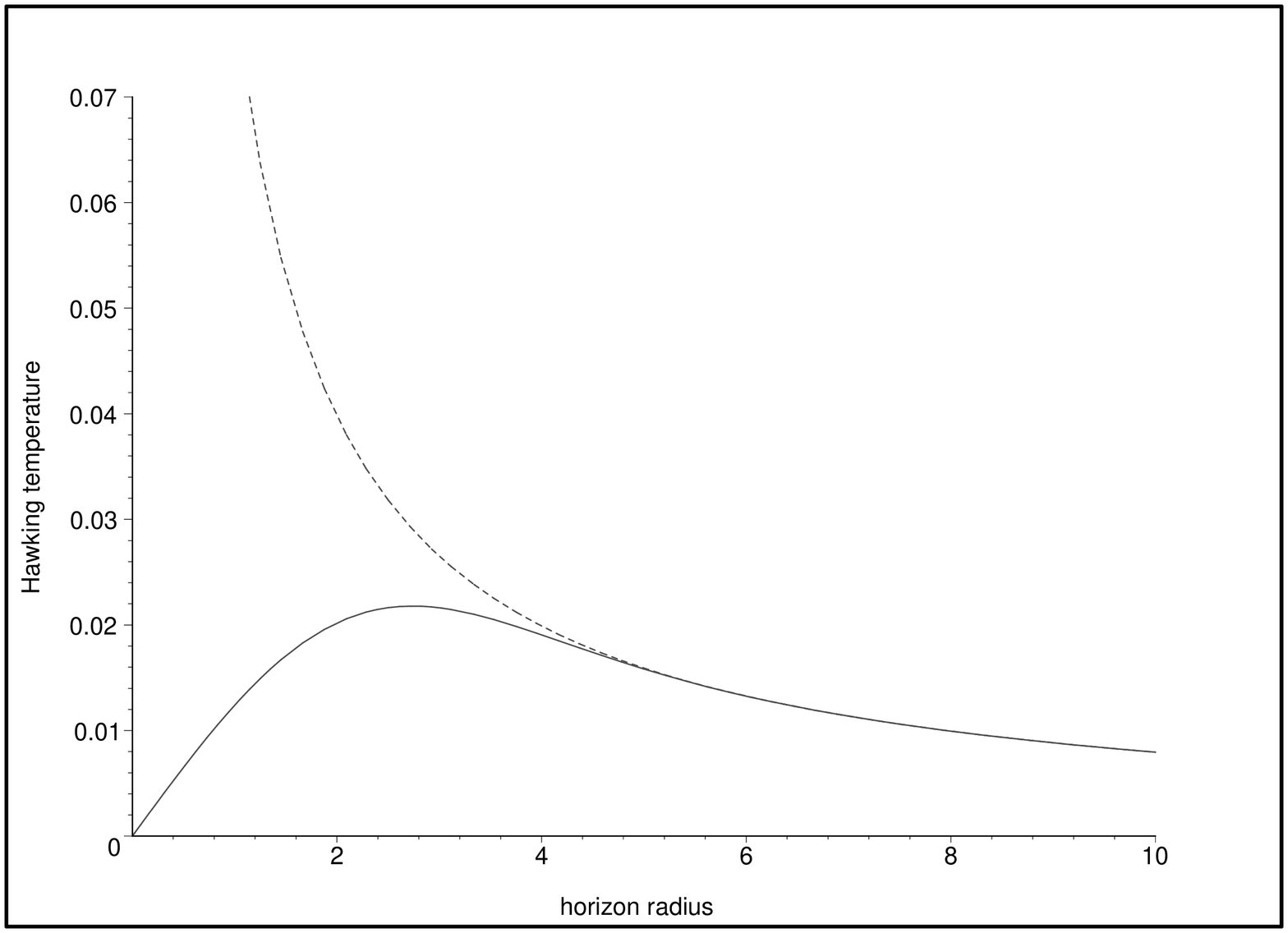}
\end{center}
\caption{Hawking temperature $T_H$ as a function of the horizon radius $r_H$%
. In the noncommutative case, solid curve, one see the temperature reaches a
maximum value $T_H^{Max.}= 2.18\times 10^{-2}/\protect\sqrt{\protect\theta}$
for $r_H= 2.74\protect\sqrt\protect\theta$, and then decreases to zero as $%
r_H\to 0$. The commutative, divergent behavior, dashed curve, is cured.}
\label{fig2}
\end{figure}
In the large radii limit, i.e. $r_H^2/4\theta>>1$, one recovers the standard
result for the Hawking temperature
\begin{equation}
T_H= \frac{1}{4\pi\, r_H}  \label{th}
\end{equation}
while noncommutativity becomes crucial, when $r_H\sim \sqrt{\theta}$. In the
conventional (~commutative~) case $T_H$ diverges and this puts limit on the
validity of the conventional description of Hawking radiation. Against this
scenario, formula (\ref{thnc}) leads to
\begin{equation}
T_H\sim \frac{r_H}{24\pi\theta} \ ,\qquad \hbox{as}\qquad \frac{r_H}{\sqrt{%
\theta}} \to 0  \label{zero}
\end{equation}
This is another intriguing result that has two important consequences.
Firstly, the emerging picture is that the black hole has reached zero
temperature and the horizon has completely evaporated. Nevertheless, we are
left with a frozen, massive, remnant. Secondly, passing from the regime of
large radius to the regime of small radius, (\ref{th}) and (\ref{zero}),
implies the existence of a \textit{maximum temperature} which is confirmed
by the plot in Figure \ref{fig2}. The plot gives the value $%
T_H^{Max.}=2.18\times 10^{-2}/\sqrt\theta $. The temperature behavior shows
that noncommutativity plays the same role in General Relativity as in
Quantum Field Theory, i.e. removes short distance divergences. The resulting
picture of black hole behavior goes as follows. For $M >> M_0$ the
temperature is given by (\ref{th}) up to exponentially small corrections,
and it increases, as the mass is radiated away. $T_H$ reaches a maximum
value at $r_H=2.74\, \sqrt\theta$, and then decreases down to zero as $r_H$
goes to zero.

At this point, important issue of Hawking radiation back-reaction should be
discussed. In commutative case one expects relevant back-reaction effects
during the terminal stage of evaporation because of huge increase of
temperature \cite{backr}. In our case, the role of noncommutativity is to
cool down the black hole in the final stage. As a consequence, there is a
suppression of quantum back-reaction since the black hole emits less and
less energy. Eventually, back-reaction may be important during the maximum
temperature phase. In order to estimate its importance in this region, let
us look at the thermal energy $E=T$ and the total mass $M$ near $%
r_H=2.74\,\sqrt\theta $. From (\ref{mh}) one finds $M\sim \sqrt\theta \,
M_{Pl.}^2$. In order to have significant back-reaction effect $T_H^{Max}$
should be of the same order of magnitude as $M$. This condition leads to the
estimate
\begin{equation}
\sqrt{\theta}\sim 10^{-1}\, l_{Pl.}\sim 10^{-34}\, cm  \label{stima}
\end{equation}
Expected values of $\sqrt{\theta}$ are above the Planck length $l_{Pl.}$ and
(\ref{stima}) indicates that back-reaction effects are suppressed even at $%
T_H^{Max}\approx 10^{18}\, GeV$. For this reason we can safely use
unmodified form of the metric (\ref{ncs}) during all the evaporation process.

As it appears, at the final stage of evaporation a mass $M_{0}$ is left
behind. One would be tempted to say that the black hole evaporation has
produced a \textit{naked singularity} of mass $M_{0}$. We are going to show
that this is not the case. In the linearized geometry the curvature tensors
can be written in terms of the Ricci scalar $R$. Thus for the metric (\ref
{ncs}) the curvature tensors are everywhere regular, since the scalar $R$
turns out to be
\begin{equation}
R=-8\pi G_{N}\rho _{\theta }  \label{ricci}
\end{equation}
\begin{figure}[h]
\begin{center}
\includegraphics[width=8cm,angle=270]{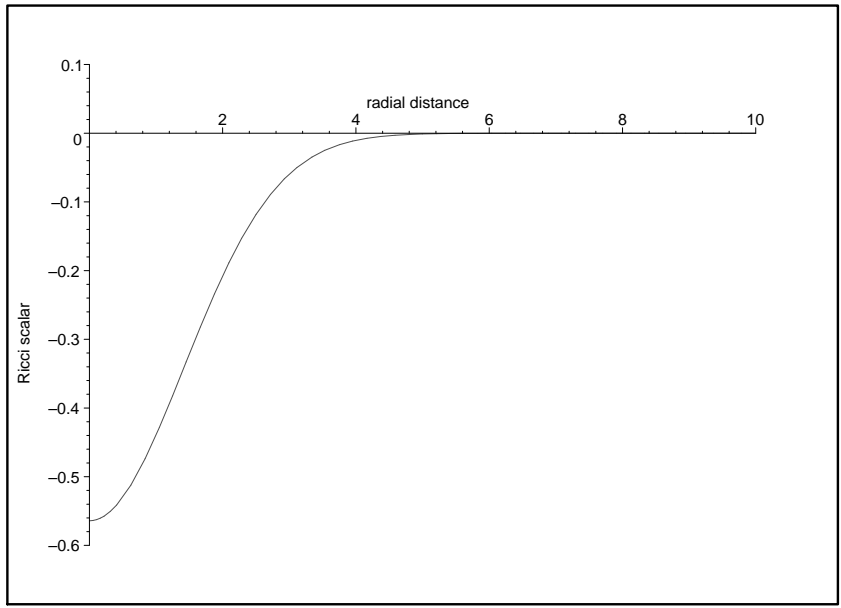}
\end{center}
\caption{Ricci scalar as function of $r$. The curvature singularity in the
origin is removed, and $R\left( \,0\,\right) =-M/\protect\sqrt{\protect\pi }%
\protect\theta ^{3/2}$.}
\label{fig3}
\end{figure}
In particular, for what concerns the case of a naked singularity one should
obtain a divergent curvature in the origin, while the short distance
behavior of $R$ is
\begin{equation}
R\simeq -\frac{M}{\sqrt{\pi }\theta ^{3/2}}  \label{ricci0}
\end{equation}
Indeed for $r<<\sqrt{\theta }$ the geometry of the frozen relic has \textit{%
constant and negative} curvature. On the other hand, in the commutative
limit $r/\sqrt{\theta }\gg 1$, one can check that (\ref{ricci}) reproduces
the usual Schwarzschild scalar curvature. Indeed the Dirac delta
distribution is restored in (\ref{ricci}) at large distances.

Regular black holes have been introduced as \textit{ad hoc} models
implementing the idea of a maximum curvature \cite{regbh}. On the other hand
we have found here an equivalent non singular black hole as a solution of
linearized Einstein equation with a source suitably prescribed by coordinate
commutativity.

As a conclusion, the results derived in this work show that the coordinate
coherent state approach to noncommutative effects can cure the singularity
problems at the terminal stage of black hole evaporation. We have shown that
noncommutativity is an intrinsic property of the manifold itself and thus
unaffected by the distribution of matter. Matter curves a noncommutative
manifold in the same way as it curves a commutative one, but cannot produce
singular structures. Specifically, we have shown that there is a minimal
mass $M_{0}=0.5\,\sqrt{\pi \theta }$ to which a black hole can decay through
Hawking radiation. The reason why it does not end-up into a naked
singularity is due to the finiteness of the curvature at the origin. The
everywhere regular geometry and the residual mass $M_{0}$ are both
manifestations of the Gaussian de-localization of the source in the
noncommutative spacetime. On the thermodynamic side, the same kind of
regularization takes place eliminating the divergent behavior of Hawking
temperature. As a consequence there is a maximum temperature that the black
hole can reach before cooling down to absolute zero. As already anticipated
in the introduction, noncommutativity regularizes divergent quantities in
the final stage of black hole evaporation in the same way it cured UV
infinities in noncommutative quantum field theory. We have also estimated
that back-reaction does not modify the original metric in a significant
manner.

\subsection*{Acknowledgement}

\textit{The author thanks the \textquotedblleft Dipartimento di Fisica
Teorica dell'Universit\`{a} di Trieste\textquotedblright , the PRIN-COFIN
project 2004 \textquotedblleft Metodi matematici per le teorie
cinetiche\textquotedblright\ and the CNR-NATO program for financial support.}

\end{document}